\documentclass[a4paper,fleqn,usenatbib]{mnras}

\usepackage{newtxtext,newtxmath}

\usepackage[T1]{fontenc}
\usepackage{ae,aecompl}

\usepackage{graphicx}	
\usepackage{amsmath}	
\usepackage{amssymb}	

\title[Spectral Puzzle of Off-Axis GRB in GW170817]
      {Spectral Puzzle of the Off-Axis Gamma-Ray Burst in GW170817}

\author[K. Ioka and T. Nakamura]{
Kunihito Ioka$^{1}$\thanks{E-mail: kunihito.ioka@yukawa.kyoto-u.ac.jp (KI)}
and Takashi Nakamura,$^{2}$
\\
$^{1}$Center for Gravitational Physics, Yukawa Institute for Theoretical Physics, Kyoto University, Kyoto 606-8502, Japan\\
$^{2}$Department of Physics, Kyoto University, Kyoto 606-8502, Japan\\
}

\date{Accepted XXX. Received YYY; in original form ZZZ}

\pubyear{2019}

\begin{document}
\label{firstpage}
\pagerange{\pageref{firstpage}--\pageref{lastpage}}
\maketitle

\begin{abstract}
  Gravitational waves from a merger of two neutron stars (NSs) were
  discovered for the first time in GW170817,
  together with diverse electromagnetic counterparts,
  providing a direct clue to the origin of short gamma-ray bursts (sGRBs).
  The associated sGRB 170817A was much fainter than typical,
  suggesting off-axis emission from a relativistic jet.
  However the observed prompt spectrum is
  inconsistent with the spectral (Amati) relation
  and causes the compactness problem
  in the simplest off-axis model.
  We focus on this spectral puzzle 
  and suggest a possible key that
  the off-axis emission generally comes from the off-center jet,
  neither the jet core nor the line-of-sight jet,
  if the jet structure is exponentially faint outward
  as inferred from observations.
  The off-center jet could be loaded with baryon or cocoon.
  The off-axis model predicts that roughly $\sim 10\%$ events
  are brighter at smaller viewing angles
  than sGRB 170817A, although the exact event rate sensitively depends on
  uncertainties of the off-center structure.
  The model also predicts outliers to Amati relation,
  providing future tests to reveal the central engine activities.
\end{abstract}




\section{Introduction}

What is the origin of short gamma-ray bursts?
This is a long-standing problem for more than 40 years
\citep[e.g.,][]{Nakar07,Berger14}.
A merger of binary neutron stars has been thought to be
the most promising candidate
\citep[e.g.,][]{Paczynski86,Goodman86,Eichler+89},
although an unequivocal evidence was missing.
The detection of the gravitational wave (GW) event GW170817 \citep{GW170817}
and the associated electromagnetic counterparts \citep{GWEM170817}
revolutionized the situation.
In particular, the short gamma-ray burst sGRB 170817A
detected two seconds ($\sim 1.7$ s) after GW170817
\citep{GRB170817A,GRB170817A_GBM,GRB170817A_INT}
and the following afterglows in radio to X-ray
\citep[][]{Troja+17,Margutti+17,Haggard+17,Hallinan+17,Alexander+17,Lyman+18}
give the first direct clues to this problem.

The problem has not been solved yet
because sGRB 170817A was very weak
with an isotropic-equivalent energy
$E_{\gamma,\rm iso} \sim 5.35 \times 10^{46}$ erg,
which is many orders of magnitude smaller than ordinary values.
On the other hand,
the afterglow observations, in particular
of superluminal motion in radio \citep{Mooley+18b,Ghirlanda+18}
and the consistency between the spectral index
and the light curve slope after the luminosity peak
\citep{Troja+18b,Mooley+18c,Lamb+19},
strongly suggest that a relativistic jet is launched and
successfully breaks out the merger ejecta in this event
\citep{Nagakura+14,Murguia-Berthier+14}.

\citet{IN18} proposed that sGRB 170817A is faint because
the jet is off-axis to our line-of-sight
\citep[see also][]{GRB170817A,Granot+17,Granot+18,LK18}.
Most emission is beamed into the on-axis direction
via a relativistic effect
and an off-axis observer receives photons emitted outside the beaming cone.
Consequently the apparent energy of the off-axis jet becomes faint
\citep{IN01,Yamazaki+02,Yamazaki+18}.
The off-axis model is initially studied by using a top-hat jet
with uniform brightness and a sharp edge.
This zeroth-order approximation is useful to capture the essential features
of the off-axis model.
Taking the effect of a finite opening angle of the jet,
a typical jet with a typical viewing angle is found to be broadly
consistent with the observations.

However the simplest off-axis model seems to be difficult to explain
the spectral peak energy $\nu_{\rm peak} = 185 \pm 62$ keV
of the main pulse of GRB 170817A,
if the central value is adopted.
The de-beamed emission from an off-axis top-hat jet tends to have
a low peak energy $\nu_{\rm peak} \sim 10$ keV,
or the peak energy viewed on-axis is required to be very high $>10$ MeV
\citep{Kasliwal+17,IN18,Matsumoto+19}.
More precisely, if the jet is viewed on-axis,
the isotropic energy and the peak energy do not satisfy
the well-known relation, so-called Amati relation for sGRBs
\citep{Amati+02,Yonetoku+04,Tsutsui+13},
which is $\sim 100$ times dimmer than that for long GRBs
at the same peak energy $\nu_{\rm peak}$.
Furthermore the afterglow observations including
VLBI observations of superluminal motion
reveal a jet with energy $E_{\rm iso} > 10^{52}$ erg,
a narrow core $\theta_c \lesssim 5^{\circ}$ and
a viewing angle $\sim 14^{\circ}$--$28^{\circ}$
\citep{Mooley+18b,Ghirlanda+18}.
If we adopt these parameters for an off-axis top-hat jet,
the compactness problem is serious \citep{Matsumoto+19}.\footnote{
  The compactness problem is in principle avoidable
  if the spectral cutoff above the peak energy is extremely sharp,
  which is not known from the observations of sGRB 170817A \citep{IN18}.
  }
In the previous paper, we put the above spectral puzzle aside
partly because the error bar is large and
the puzzle is solved if we allow 3$\sigma$ error \citep{IN18}.
However the large error bar could mainly stem from
the spectral evolution during the burst
and the peak energy of the main pulse could be really high \citep{Veres+18}.
The another reason for neglecting the spectral puzzle is that the weak tail
with 34\% of the fluence of the main pulse
has a temperature $k_B T = 10.3\pm 1.5$ keV,
consistent with the simplest off-axis model.
The main pulse could be produced by a different mechanism,
such as scattering of an sGRB by a cocoon \citep{Kisaka+15,Kisaka+17,Kisaka+18}
and a cocoon breakout from ejecta
\citep{Kasliwal+17,Gottlieb+18,Nakar+18,KI+19}.

The approximation of a top-hat jet would be too simple
to discuss the spectral puzzle \citep{Meszaros+98,ZM02}.
The afterglow observation,
in particular the slowly-rising light curve,
is not consistent with a top-hat jet \citep{Mooley+18a},
but strongly suggests a structured jet
\citep{Troja+18,Ruan+18,Margutti+18,D'Avanzo+18,Lazzati+18,Lyman+18,Troja+18b,Ghirlanda+18,Lamb+19}.
The jet structure most likely affects
the prompt emission and its spectrum as well.

In this paper we reexamine
the spectrum of an off-axis jet in GW170817
taking the jet structure constrained by the afterglow observations into account.
We suggest a possible solution, in which
the off-axis emission from a typical sGRB jet
is compatible with sGRB 170817A.
In Sec.~\ref{sec:structure},
we summarize the jet structure inferred from the afterglow observations.
In Sec.~\ref{sec:formulation},
we generalize the formulation in \citet{IN01} to calculate
the off-axis emission from a structured jet.
In Sec.~\ref{sec:off-center},
we show that the off-axis emission mostly comes from the off-center jet,
neither the jet core nor the line-of-sight jet,
if the jet structure is exponentially faint outward,
and the off-center emission can solve the spectral puzzle.
Section~\ref{sec:summary} is devoted to summary and discussions.

\section{Jet structure from afterglow observations}\label{sec:structure}

\begin{figure}
  \begin{center}
    \includegraphics[width=\linewidth]{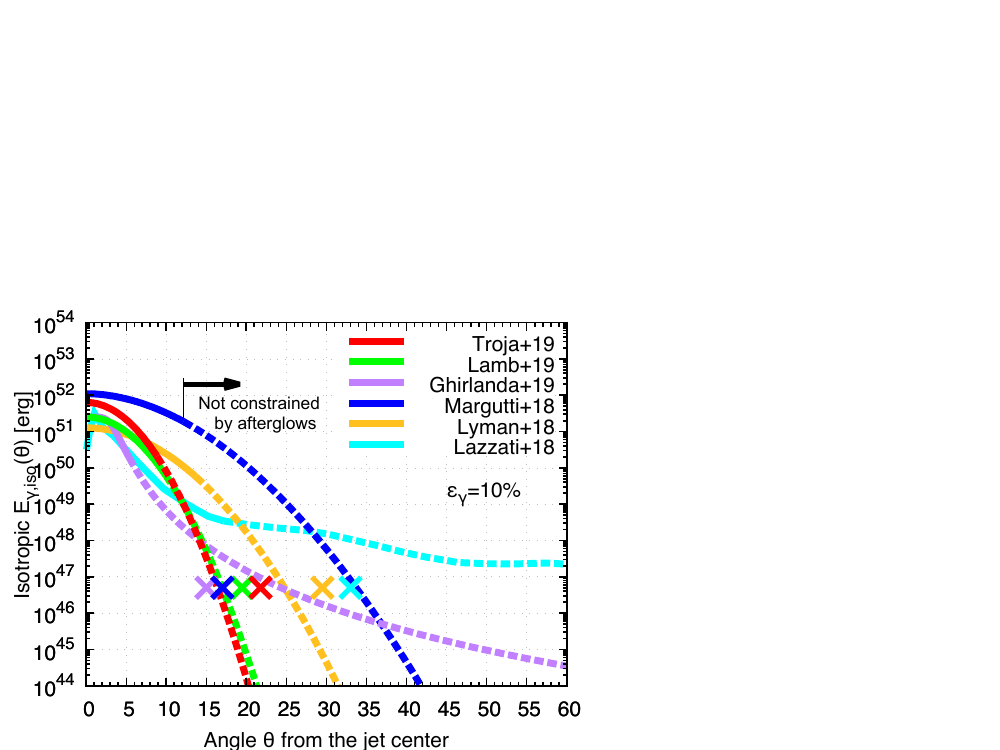}
  \end{center}
  \caption{
    The jet structures
    inferred from the observations of the afterglow light curves
    ({\it solid lines}) are plotted,
    where a radiative efficiency $\epsilon_{\gamma}=10\%$ is assumed.
    The {\it dashed lines} show the part unconstrained
    by the afterglow observations.
    The isotropic energy of sGRB 170817A is also plotted
    at the viewing angles of the jet ({\it cross marks}).
    See the text for details.
  }
  \label{fig:profile}
\end{figure}

Before discussing the off-axis emission,
we summarize the jet structure
inferred from the observations of the afterglows
in Fig.~\ref{fig:profile}. The solid and dashed   lines show
the gamma-ray energy of each model specified by different color
as a function of an angle $\theta$ from the jet center.
The afterglow of sGRB 170817A
was observed after $\sim$ 10 days from the prompt sGRB,
and the solid line part of the jet emission was not observed at $\sim$ 10 days 
since only the part with $\theta >  \theta_v-\Gamma^{-1}$ has
a significant contribution to the afterglow radiation, 
where $\Gamma$ and $\theta_v$ are the Lorentz factor of the emitting region
and the viewing angle, respectively.
Here we obtain  $\Gamma$
assuming that each fluid element at each angle follows
the spherical hydrodynamical evolution decelerated in the interstellar medium
because numerical simulations suggest that this is a good approximation
\citep{KG03,vanEerten+10}.
Note that the Lorentz factor $\Gamma$ in the afterglow phase is
basically different from that in the prompt phase. 

As time goes, the angular size $\Gamma^{-1}$ expands.
Then we can progressively probe the jet structure outside
the angular size observable at $\sim 10$ days,
in particular the inner part close to the jet axis
because the inner jet dominates the energy and the flux.
However we can not infer the structure of the dashed region
because the emission from this region is integrated in the initial observation
at $\sim$ 10 days.
Thus the dashed lines are
the part unconstrained by the afterglow observations.
The observations of the afterglow start at $\sim 10$ days
and there is no information before that.
At this time, $\sim 10$ days, we are already observing
an angular size of $\sim 1/\Gamma$ around the line-of-sight in the jet.

In Fig.~\ref{fig:profile},
we assume a radiative efficiency $\epsilon_{\gamma}=10\%$,
which is typical for the observed sGRBs \citep[e.g.,][]{Fong+15}.
These structures are obtained
by assuming a functional form of the jet structure,
calculating the afterglow emission with the standard afterglow theory,
and fitting the afterglow observations
with model parameters such as the ambient density $n$,
the viewing angle of the jet $\theta_v$, 
the total energy, and so on.
The different structures obtained by different authors
most likely reflect the difference of the assumed structure
and model parameters,
which are not completely determined
only by the light curve observations.

In Fig.~\ref{fig:profile}, we also plot
the isotropic energy of sGRB 170817A
at the viewing angle of the jet ({\it cross marks})
to compare it with the jet structure.
Note that the VLBI observations of superluminal motion
prefer a viewing angle of $14^{\circ} \lesssim \theta_v \lesssim 28^{\circ}$
\citep{Mooley+18b}.
Note also that in blue, cyan and purple line cases,
the gamma-ray energy exceeds
that of sGRB 170817A,
requiring smaller radiative efficiency than $\epsilon_{\gamma}=10\%$
at the viewing angle.

From Fig.~\ref{fig:profile},
we can find that 
the central part should be much more energetic than
the observed sGRB 170817A,
regardless of the different structures obtained by the different authors.
In order not to exceed sGRB 170817A,
the isotropic gamma-ray energy of the jet
should decrease exponentially outward
(where it is not always Gaussian but
a sharply decreasing function).
This is a general property required from the afterglow and sGRB 170817A.
Therefore we adopt a fiducial case as
\begin{eqnarray}
  E_{\gamma}(\theta)=\epsilon_{\gamma} E_0 \exp(-\theta^2/2\theta_c^2),
  \label{eq:stjet}
\end{eqnarray}
with $E_0=10^{52.80}$ erg,
$\theta_c=0.059$,
$n=10^{-2.51}$ cm$^{-3}$,
and $\theta_v=0.38 \approx 22^{\circ}$ \citep{Troja+18b}.
Note that although we do not know whether the structure
reflects that of the jet energy or of the radiative efficiency,
it does not matter to the following discussions.
Note also that although the jet structure could be modified
after the prompt emission,
namely during the propagation in the interstellar medium,
it does not change the above conclusion that
the jet structure is exponentially faint outward.

\section{Formulation of off-axis emission}\label{sec:formulation}

To calculate the off-axis emission from a structured jet,
we generalize the formulation in \citet{IN01} in this section.
We consider an axisymmetric jet for simplicity.
Assuming that the distance to the source $d$
is much larger than the source size,
the observed flux $F_{\nu}$ at a frequency $\nu$
is obtained from volume integration
of the emission coefficient $j_{\nu}$ as
\begin{eqnarray}
  F_{\nu}
  &\simeq&\frac{1}{d^2} \int r^2 dr \sin\theta\, d\theta\, d\phi\, j_{\nu},
  \label{eq:Fnu}
\end{eqnarray}
where the jet has an origin at $r=0$ and an axis at $\theta=0$ in 
the spherical coordinate $(r, \theta, \phi)$.
The jet axis has a viewing angle $\theta_v$
from the line-of-sight between the observer and the origin.

The Lorentz transformation of the emission coefficient and frequency
from the lab frame (i.e., source center frame) $j_{\nu}$
to the comoving frame $j'_{\nu'}$ is
\begin{eqnarray}
  j_{\nu}&=&\frac{j'_{\nu'}}{\Gamma^2(1-\beta \cos\theta_{\Delta})^2},
  \label{eq:jnu}
  \\
  \nu&=&\frac{\nu'}{\Gamma (1-\beta \cos\theta_{\Delta})},
\end{eqnarray}
respectively
where we assume that the jet moves in the radial direction and thereby
the angle $\theta_\Delta$ between the jet motion
and the line-of-sight direction is given by
that between the $(\theta,\phi)$ direction and
the line-of-sight direction as
\begin{eqnarray}
  \cos\theta_{\Delta}=\sin\theta \cos\phi \sin\theta_v + \cos\theta \cos\theta_v.
\end{eqnarray}

A single pulse of sGRBs is well approximated by 
instantaneous thin-shell emission at time $t_0(\theta)$ and radius $r_0(\theta)$,
\begin{eqnarray}
  j'_{\nu'}=\frac{1}{(4\pi)^2 r^2} E'_{\gamma}(\theta)
  f(\nu',\theta) \delta[r-r_0(\theta)] \delta[t-t_0(\theta)],
  \label{eq:jnu_model}
\end{eqnarray}
where the angular structure of the jet is characterized by
the comoving radiation energy $E'_{\gamma}(\theta)$ [erg].
This is related with the radiation energy $E_{\gamma}(\theta)$
and total energy $E(\theta)$ in the lab frame as
\begin{eqnarray}
  \epsilon_{\gamma} E(\theta)
  =E_{\gamma}(\theta)
  =\Gamma E'_{\gamma}(\theta),
  \label{eq:E'}
\end{eqnarray}
where the Lorentz factor $\Gamma$
and the radiative efficiency $\epsilon_{\gamma}$
also have angular structures in general.
We adopt the spectral shape similar to the so-called Band function
\begin{eqnarray}
  f(\nu',\theta) = \frac{C}{\nu'_0(\theta)}
  \left(\frac{\nu'}{\nu'_0(\theta)}\right)^{1+\alpha_B}
  \left[1+\left(\frac{\nu'}{\nu'_0(\theta)}\right)^2\right]^{\frac{\beta_B-\alpha_B}{2}},
  \label{eq:f(nu',theta)}
\end{eqnarray}
with $\alpha_B\sim -1$ and $\beta_B\sim -2.5$
\citep{Kaneko+06}.
We take the constant $C$ so that $\int d\nu' f(\nu',\theta)=1$.
Note that the following discussions do not depend on
the exact shape of the spectrum as long as it has a peak.

The time in the lab frame $t$ is related with
the observed time $T$ as
\begin{eqnarray}
  t=T+\frac{r}{c}\cos\theta_{\Delta},
  \label{eq:times}
\end{eqnarray}
where the time is measured from the merger time and we neglect the cosmological effect.

The isotropic energy is obtained from Eqs.~(\ref{eq:Fnu}), (\ref{eq:jnu}), (\ref{eq:jnu_model}), (\ref{eq:E'}) and (\ref{eq:times})
by performing the integrals of the delta functions as
\begin{eqnarray}
  E_{\gamma,\rm iso}&=&\int dT \int d\nu\ 4\pi d^2 F_{\nu}
  \nonumber\\
  &=& \frac{1}{4\pi} \int \sin\theta\, d\theta\, d\phi\,
  \frac{E_{\gamma}(\theta)}{\Gamma^4(1-\beta \cos\theta_{\Delta})^3},
  \label{eq:Eiso1}
\end{eqnarray}
where the arbitrary functions $r_0(\theta)$ and $t_0(\theta)$
are integrated out.
We can further perform the $\phi$ integral,
\begin{eqnarray}
  E_{\gamma,\rm iso}=\int \frac{\sin\theta\, d\theta}{2}\,
  E_{\gamma}(\theta)\cdot \mathcal{B}(\theta),
  \label{eq:Eiso2}
\end{eqnarray}
where we call the last part as the beaming term,
\begin{eqnarray}
  \mathcal{B}(\theta)&\equiv &\int_{-\pi}^{\pi} \frac{d\phi}{2\pi}\,
  \frac{1}{\Gamma^4(1-\beta \cos\theta_{\Delta})^3}
  \nonumber\\
  &=&\frac{1}{2\Gamma^4}
  \frac{2\left(1-\beta \cos\theta \cos\theta_v\right)^2
    +\left(\beta \sin\theta \sin\theta_v\right)^2}
       {[1-\beta\cos(\theta_v+\theta)]^{5/2}
         [1-\beta\cos(\theta_v-\theta)]^{5/2}}.
       \label{eq:beam}
\end{eqnarray}
Note that we can explicitly show $E_{\gamma,\rm iso}=E_{\gamma}(\theta)$
if $E_{\gamma}(\theta)$ and $\Gamma(\theta)$ are isotropic
(where we can always put $\theta_v=0$ by changing a coordinate
in the integration).

The surface brightness (i.e., the isotropic energy per solid angle) is given by
\begin{eqnarray}
  \frac{dE_{\gamma,\rm iso}}{d\Omega}=
  \frac{1}{4\pi}\frac{E_{\gamma}(\theta)}{\Gamma^4(1-\beta \cos\theta_{\Delta})^3}.
  \label{eq:surface}
\end{eqnarray}

The spectral peak energy $\nu_{\rm peak}$ corresponds to the energy
at which $\nu dE_{\gamma,\rm iso}/d\nu$ takes a maximum value.
We can show
\begin{eqnarray}
  \frac{dE_{\gamma,\rm iso}}{d\nu}
  = \frac{1}{4\pi} \int \sin\theta\, d\theta\, d\phi\,
  \frac{E_{\gamma}(\theta) f(\nu,\theta,\phi)}{\Gamma^4(1-\beta \cos\theta_{\Delta})^3},
  \label{eq:spec}
\end{eqnarray}
where
\begin{eqnarray}
  f(\nu,\theta,\phi) = \frac{C}{\nu_0(\theta,\phi)}
  \left(\frac{\nu}{\nu_0(\theta,\phi)}\right)^{1+\alpha_B}
  \left[1+\left(\frac{\nu}{\nu_0(\theta,\phi)}\right)^2\right]^{\frac{\beta_B-\alpha_B}{2}},
\end{eqnarray}
and $\nu_0(\theta,\phi)=\nu'_0(\theta)/\Gamma(1-\beta\cos\theta_{\Delta})$.

\section{Off-axis emission comes from off-center}\label{sec:off-center}

\subsection{Dominance of off-axis emission}

\begin{figure}
  \begin{center}
    \includegraphics[width=\linewidth]{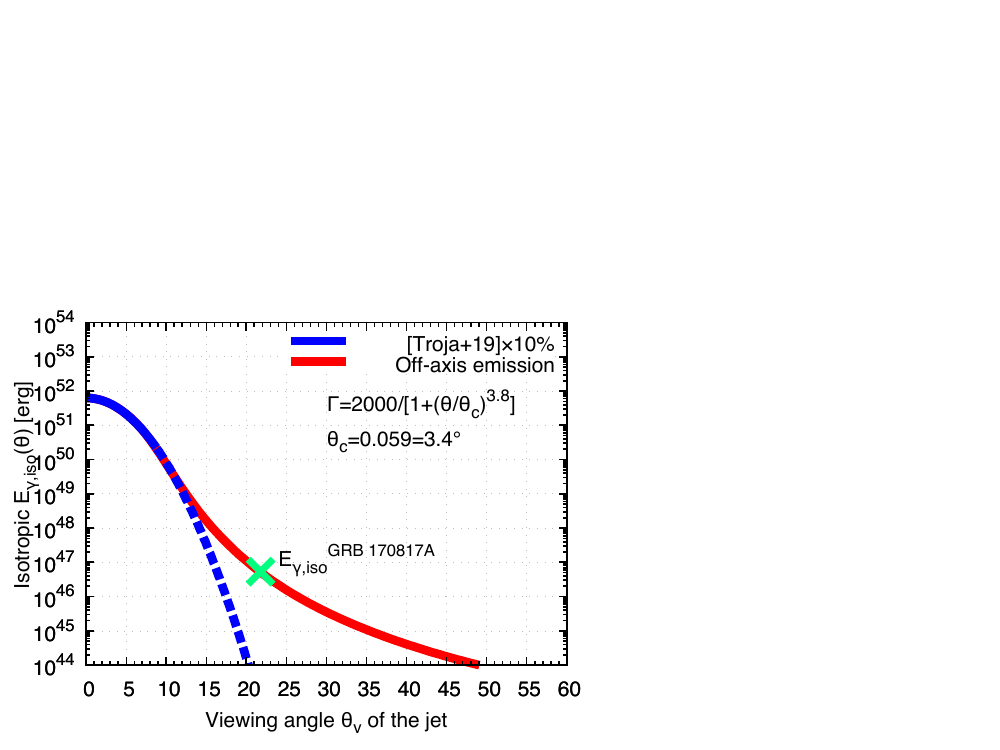}
  \end{center}
  \caption{
    The isotropic gamma-ray energy of
    the off-axis emission ({\it red line}) from
    a structured jet in Eq.~(\ref{eq:stjet}) ({\it blue line})
    is plotted as a function of the viewing angle $\theta_v$ of the jet.
    The isotropic gamma-ray energy at the viewing angle of sGRB 170817A
    is also plotted ({\it green cross}).
    We adopt the Lorentz factor profile in Eq.~(\ref{eq:gamma}).
    The off-axis emission always dominates
    the line-of-sight emission in the outer region.
  }
  \label{fig:Eiso}
\end{figure}

In Fig.~\ref{fig:Eiso},
we calculate the off-axis emission ({\it red line}) from
a structured jet in Eq.~(\ref{eq:stjet}) ({\it blue line})
with Eqs.~(\ref{eq:Eiso2}) and (\ref{eq:beam}).
For the calculation we need the Lorentz factor,
which is not well constrained from observations.
As a fiducial Lorentz factor,
we adopt a profile decreasing outward
\begin{eqnarray}
  \Gamma=\frac{\Gamma_{\max}}{1+(\theta/\theta_c)^{\lambda}}.
  \label{eq:gamma}
\end{eqnarray}
We take $\Gamma_{\max}=2000$
since lower limits $\Gamma \gtrsim 1000$
are obtained for some sGRBs like sGRB 090510 detected by {\it Fermi}/LAT.
The index $\lambda$ is used
to match the isotropic energy with that of sGRB 170817A,
and is found to be $\lambda \approx 3.8$
for our fiducial case in Eq.~(\ref{eq:stjet}).
It is always possible to match the observed value.
Note that $\Gamma>1$ for our range of interest.
Even if the Lorentz factor profile is different,
the following discussions are similar as long as
it is smooth enough.

As shown in Fig.~\ref{fig:Eiso},
the off-axis emission ({\it red line})
always dominates the line-of-sight emission ({\it blue line})
in the outer region.
This is general
irrespective of the uncertainty of the jet structure for GW170817
because the jet energy should decrease exponentially outward
in order to satisfy both the afterglow observation
(i.e., the large energy at $\theta=0$)
and the prompt sGRB observation (i.e., the small energy at $\theta=\theta_v$)
while the off-axis emission has a power-law profile
$\propto (\theta_v-\theta_c)^{-4}$
\citep{IN18}.
Therefore if sGRB 170817A arises from a jet,
it is most likely off-axis emission, not the line-of-sight emission.

\subsection{Off-center emission}

\begin{figure}
  \begin{center}
    \includegraphics[width=\linewidth]{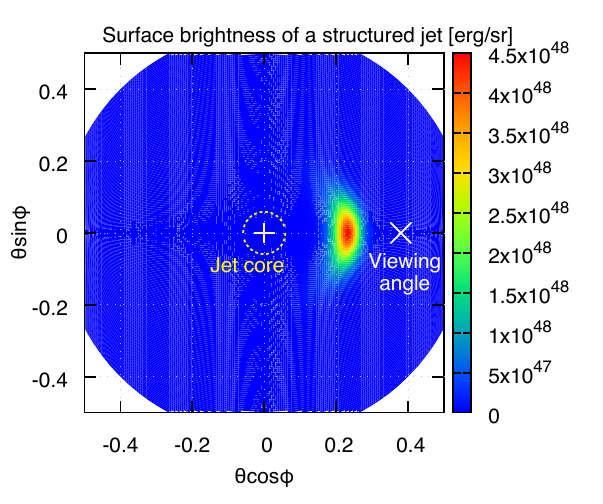}
  \end{center}
  \caption{
    The surface brightness distribution of the jet emission is plotted
    on the $(\theta \cos\phi, \theta \sin\phi)$ plane.
    The model parameters are the same as in Fig.~\ref{fig:Eiso}.
    Most emission comes from the off-center jet,
    neither the jet core nor the line-of-sight jet but the middle.
  }
  \label{fig:surface}
\end{figure}

Which part of the jet makes a major contribution
to the observed off-axis emission?
In Fig.~\ref{fig:surface}, we plot
the surface brightness distribution in Eq.~(\ref{eq:surface})
with the same model parameters as in Fig.~\ref{fig:Eiso}.
It is remarkable that most emission comes from the off-center jet,
neither the jet core nor the line-of-sight jet but the middle.

\begin{figure}
  \begin{center}
    \includegraphics[width=\linewidth]{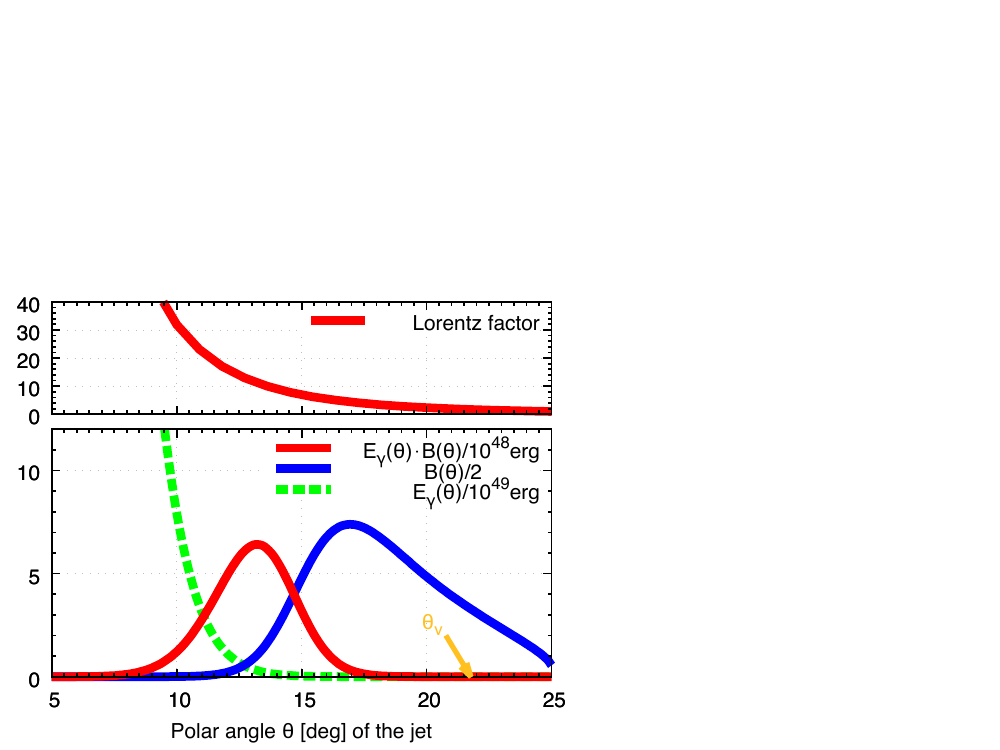}
  \end{center}
  \caption{
    ({\it Lower panel}): The jet energy ($E_{\gamma}(\theta)$; {\it green line}),
    the beaming term in Eq.~(\ref{eq:beam})
    ($\mathcal{B}(\theta)$; {\it blue line})
    and their product ({\it red line}), which determines
    the isotropic energy in Eq.~(\ref{eq:Eiso1}),
    are plotted as a function of the polar angle $\theta$ of the jet.
    The model parameters are the same as in Fig.~\ref{fig:Eiso}.
    The product of the decreasing $E_{\gamma}(\theta)$ and
    the increasing $\mathcal{B}(\theta)$
    makes a peak in the off-center region,
    neither the jet core ($\theta_c=0.059=3.4^{\circ}$)
    nor the line-of-sight jet ($\theta_v=0.38$; {\it orange arrow}).
    ({\it Upper panel}): The Lorentz factor distribution
    in Eq.~(\ref{eq:gamma}) is plotted as a function of $\theta$.
  }
  \label{fig:LF}
\end{figure}

The off-center emission is also
a general property of the off-axis emission
irrespective of the uncertainty of the adopted jet structure for GW170817.
As we can see from Eq.~(\ref{eq:Eiso2}),
the observed isotropic energy is determined by the product
of the jet structure $E_{\gamma}(\theta)$ and
the beaming term $\mathcal{B}(\theta)$.
These two functions are plotted in Fig.~\ref{fig:LF} ({\it lower panel}).
The jet energy ({\it green line}) should decrease exponentially outward
to satisfy both the observations of the afterglow and sGRB 170817A.
On the other hand the beaming term ({\it blue line}) increases outward
within $\theta \lesssim \theta_v - \Gamma^{-1}$
because the beaming cone approaches the line-of-sight.
The product of the decreasing function ($E_{\gamma}(\theta)$)
and the increasing function ($\mathcal{B}(\theta)$)
makes a peak in the middle,
neither the jet core nor the line-of-sight.\footnote{
  This is analogous to a Gamov peak.}

\subsection{Compactness problem}

The off-center emission is important
to avoid the compactness problem.
The isotropic energy of the emitting region is
$\sim 10^{48}$--$10^{49}$ erg
much smaller than that of the jet core $\sim 10^{52}$ erg.
The angular separation
between the emitting region and the line-of-sight
is also $\sim 5^{\circ}$--$10^{\circ}$ much smaller than
the top-hat case $\theta_v-\theta_c \sim 20^{\circ}$.
Taking into account that
the Lorentz factor at the emitting region $\Gamma \sim 10$ 
(see {\it upper panel} in Fig.~\ref{fig:LF})
is smaller than that of the jet core $\Gamma \sim 10^3$,
we can show that the compactness problem is not crucial
\citep{Matsumoto+19}.\footnote{
  The innermost part of the emitting region
  could suffer from the compactness problem.
}

\subsection{Spectral Amati relation}

The off-center emission could also solve
the seemingly inconsistency with the Amati relation
between the spectral peak energy and the isotropic energy.
Actually the off-center emission does not satisfy the Amati relation
as argued previously.
However the jet core can satisfy the Amati relation.
The point is that the off-center jet is different from the jet core.
The classical sGRBs are emitted by the on-axis jet core
and hence satisfy the Amati relation,
where the off-center region is too faint to observe.
The off-center region is observable only for an off-axis event
like sGRB 170817A,
in which the core emission is severely de-beamed and suppressed.
Usually an off-axis event is difficult to find
because they are faint.
The jet structure is exponentially faint outward
and is more or less top-hat-like.
Thus it is easy to understand that
the off-axis event is difficult to find
without any other signature like GWs.

\begin{figure}
  \begin{center}
    \includegraphics[width=\linewidth]{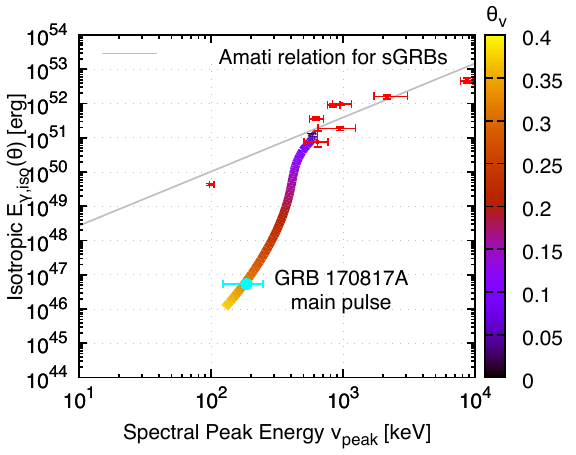}
  \end{center}
  \caption{
    An example where
    the on-axis emission from the jet core satisfies the Amati relation
    and the off-axis emission from the off-center jet reproduces sGRB 170817A
    is constructed on the plane of the spectral peak energy $\nu_{\rm peak}$
    and isotropic energy $E_{\gamma,{\rm iso}}$.
    The color of the line shows the viewing angle of the jet $\theta_v$.
    The model is compared with the observations of sGRBs, which satisfies
    the Amati relation ({\it grey solid line}).
    Note that Amati relation for sGRBs is 
    $\sim 100$ times dimmer than that for long GRBs at the same peak energy
    $\nu_{\rm peak}$,
    which is determined by the sGRBs used in \citet{Tsutsui+13}
    ({\it red filled squares}).
  }
  \label{fig:spec}
\end{figure}

In Fig.~\ref{fig:spec}, we construct a concrete example
of the off-axis emission where
the on-axis emission from the jet core satisfies the Amati relation
and the off-axis emission from the off-center jet 
reproduces sGRB 170817A.
We adjust the comoving spectral energy $\nu'_0(\theta)$
in Eq.~(\ref{eq:f(nu',theta)})
as $\nu'_0(\theta)=0.15\, {\rm keV} [1+(\theta/\theta_c)^{3.4}]$
by changing the power-law index (3.4 here)
with the same model parameters as in Figs.~\ref{fig:Eiso}--\ref{fig:LF}.
Note that the product with Eq.~(\ref{eq:gamma})
  $\Gamma(\theta) \nu'_0(\theta)$ (i.e., lab-frame spectral energy) is
  a slowly decreasing function of $\theta$.
The Amati relation is satisfied
as long as the viewing angle is about the core size $\theta_v \sim \theta_c$,
where $\theta_c=0.059=3.4^{\circ}$ for our fiducial case.
As the viewing angle becomes large,
the spectral relation deviates from the Amati relation
and approaches the point of sGRB 170817A.
If this interpretation is correct,
future observations will find outliers of Amati relation,
although the event rate of the outliers is very small
because the jet structure is exponentially faint outward.

\section{Summary and Discussions}\label{sec:summary}

In this paper we identify the spectral puzzle
and propose a natural solution 
in the off-axis jet model for sGRB 170817A associated with GW170817.
Taking both the afterglow and prompt sGRB 170817A into account,
we first clarify that the jet structure should be exponentially faint outward,
so that the off-axis emission dominates the line-of-sight emission
at a large viewing angle.
Given that, 
we generally show that the off-axis emission comes from an off-center jet,
neither the jet core nor the line-of-sight but the middle.
Because the off-center jet is much less energetic 
and much closer to the line-of-sight than the jet core,
the compactness problem is avoidable.
In addition, the Amati relation may be violated as in sGRB 170817A
because the off-center jet is different from the jet core
that satisfies the Amati relation.

The off-center jet is too faint to find for ordinary observations
without any other messengers like GWs,
high-energy gamma-rays \citep{Murase+18} and neutrinos \citep{Kimura+18}.
Even if it is detected in gamma-rays, the redshift is not determined
without follow-up observations.
Since the jet structure is exponentially faint outward and hence top-hat-like,
the event rate of detecting the off-center jet is very small.
This explains why there are not so much outliers to the Amati relation.
Conversely the off-axis model predicts outliers to the Amati relation
once many events are observed.
It is reported that there are similar bursts with similar spectrum to
sGRB 170817A among previous sGRBs \citep{Burns+18,Troja+18a,vonKienlin+19}.
They could be off-axis events like sGRB 170817A.

The off-axis model predicts
brighter events at smaller viewing angles
than sGRB 170817A.
Taking the inclination dependence of the GW amplitude into account,
the probability of observing the viewing angle
$\theta_v \lesssim 20^{\circ}$, $15^{\circ}$ and $10^{\circ}$
is $\sim 20\%$, $10\%$ and $5\%$, respectively
\citep{LK17}.
Therefore future multi-messenger observations,
in particular with GWs and gamma-rays,
can test the off-axis jet model in GW170817.
However, the exact event rate of the off-axis event depends on
the outer structure of the jet,
which is not constrained by
the afterglow observations of GW170817
(see {\it dashed lines} in Figs.~\ref{fig:profile} and \ref{fig:Eiso}).
It is necessary to observe earlier afterglows than $\sim 10$ days
to probe the off-center jet that is relevant to the off-axis emission.
In addition the viewing angle of GW170817
still has an uncertainty $\theta_v \sim 14^{\circ}$--$28^{\circ}$
even after the VLBI observations of superluminal motion
because it depends on the jet core size and
the sideway expansion during the propagation in the interstellar medium,
which also depends on the unobserved off-center structure.
It is a future problem to encompass the uncertainty of the event rate.

It is not known how the jet structure is formed
in particular in the off-center region for sGRB 170817A.
There are several possibilities.
First the jet itself could have a structure.
Even if the jet has a weak outer structure,
it is cut off during the jet propagation through the merger ejecta.
However the baryon load could be angular dependent,
leading to a different angular expansion and hence
a jet structure after the jet breakout from the merger ejecta.
Note that baryon may be loaded at the base of the jet launch
as well as during the jet propagation.
Second the off-center region could be a cocoon,
which is produced by the shocked jet and shocked ejecta
during the jet propagation.
Although a typical velocity of the cocoon is sub-relativistic,
an imperfect mixing between the shocked jet and shocked ejecta
make a high-entropy cocoon surrounding a jet,
which could lead to a jet structure.
We did not also consider several effects such as
the temporal evolution,
the photosphere,
the high-energy spectral cutoff
and so on.
These are interesting future problems.

\section*{Acknowledgements}

The authors would like to thank
J.~Granot, H.~Hamidani, K.~Hotokezaka, K.~Kashiyama,
T. Kinugawa, S.~Kisaka, K. Kyutoku,
G.~P.~Lamb, A.~Levinson, T.~Matsumoto,
T.~Piran, G.~Ryan, M. Shibata, K.~Takahashi and R.~Yamazaki
for useful discussions.
This work is partly supported by
JSPS KAKENHI nos. 18H01215, 17H06357, 17H06362, 17H06131, 26287051
(KI) and 15H02087 (TN).




\bibliographystyle{mnras}
\bibliography{ref} 







\bsp	
\label{lastpage}
\end{document}